\documentclass[12pt]{iopart}

\usepackage{iopams}
\usepackage{graphicx,graphics,color,epsfig}
\usepackage{times}
\usepackage{txfonts}
\usepackage{epstopdf}
\usepackage{latexsym}
\usepackage{epsfig}
\usepackage{setstack}
\usepackage{bm}
\usepackage{cite}

\newcommand{\ket}[1]{| #1 \rangle}

\begin{document}

\title[Effect of retardation on the evolution of entanglement]{Effect of retardation in the atom-field interaction on entanglement in a double Jaynes-Cummings system}

\author{Smail Bougouffa$^{1}$ and Zbigniew Ficek$^{2}$}
\address{$^{1}$Department of Physics, Faculty of Science, Taibah University, P.O. Box 30002, Madinah, Saudi
Arabia\\ 
$^{2}$The National Center for Mathematics and Physics, KACST, P.O. Box 6086, Riyadh~11442, Saudi Arabia}  
\ead{sbougouffa@taibahu.edu.sa, zficek@kacst.edu.sa}

\date{\today}

\begin{abstract}
The effect of retardation in the atom-field interaction on the dynamics of entanglement in a double Jaynes-Cummings system is investigated.
We consider large cavities in which a finite time necessary for light to travel between the atoms and the cavity mirrors may result in retardation effects. Our results demonstrate the qualitatively new behaviour observable in the time evolution of entanglement when the retardation effects are included. Solutions for single and double excitation in the system are presented. We follow the temporal evolution of an initial entanglement and find that the evolution is affected drastically by the retardation effects. In particular, the harmonic oscillations of the atomic populations and the concurrence, characteristic of single-mode Jaynes-Cummings systems, are suppressed when the retardation effects are included. The process of revival of the entanglement degrades with an increasing number of the cavity modes to which the atoms are coupled. It is also found that the effect of the retardation on the doubly excited states is more drastic than on the single excitation states that at relatively short times, the retardation leads to a complete distortion of entanglement carried by a doubly excited state. 
\end{abstract}

\date{\today}
\maketitle

\section{Introduction}

Partly because of the successful experimental realisation of quantum gates, and the opportunity it provides for applications in quantum information and quantum computation, the problem of the entanglement evolution and transfer in a network of quantum systems has recently received a great deal of attention~\cite{10,11}. The theoretical description of such systems is generally complicated, in particular when one includes decoherence processes resulting from the coupling of a system to external environment. The complete treatments of these situations are available. However, the physics of the coherent processes involved in the transfer of entanglement are often lost in details or masked by the incoherent processes~\cite{12,14,15,16,18,19,20}. 

Although less important from the practical situation, studying the evolution and transfer of entanglement in simpler systems such as Jaynes-Cummings models~\cite{JC63,pa63} involving single atoms interacting with a single-mode cavity field is significant in terms of giving greater insight into the fundamentals of the transfer process. In particular, the relation between the form of entangled states and the efficiency of the transfer can be studied. A large number of studies of such simple systems have been carried out for independent~\cite{Eberly1,Eberly2,Eberly3,yelattice,sainz,Ficek1,S10,TMSZ10,SA11} and also coupled Jaynes-Cummings models~\cite{nr07,oi08,ew10,dz12,gl12,xf12,pe97,sm06,pl07,zh09,zy10,sl12}. In these models simple analytical formulas can be derived for the state vector or density operator of the system and the relations between the coherence properties and the efficiency of transferring a given state can be studied.   

It has been demonstrated that the evolution of singly and doubly excited entangled states in independent or coupled Jaynes-Cummings systems can be significantly different that the singly excited states can be continuously and completely transferred between the atoms and the cavity modes whereas doubly excited states may suffer an abrupt loss during the evolution, the phenomenon known as sudden death of entanglement~\cite{13,17,col}. Although the single-mode approximation of the interaction of an atom with a cavity field is often entirely adequate to describe a practical situation, there are circumstances when the interaction of the atom with a large number of modes cannot be neglected, for example when sizes of cavities are much larger than the wavelength of the atomic transition. In this case, a finite number of the cavity modes could fit into the atomic linewidth leading to an effective multi-mode interaction of the atom with the cavity field. The atom can then be re-excited at finite time and at multiple finite times required for a photon emitted by the atom to cycle through the cavity and be reabsorbed. The finite interaction time between the atom and the cavity field, arising from a finite time required for the photon to travel between the atom and the cavity mirrors, could lead to retardation in the excitation of the atom~\cite{42,43,44,Meyster1,gf12}. 

In this paper, we show how various results relating to the evolution of entanglement in a system composed of two isolated Jaynes-Cummings models can be affected by the retardation. We calculate concurrence, the measure of entanglement between two qubits, and investigate the effect of retardation on the evolution of single and double excitation entangled states. We find that the evolution differs qualitatively from that predicted before in the absence of the retardation. In particular, the evolution is essentially nonoscillatory. A simple physical interpretation is given based on the energy-time uncertainty relation.

The paper is organised as follows. We begin in section~\ref{sec1} by introducing the model and explaining in details the meaning of retardation in the atom-field interaction. In section~\ref{sec2} we derive equations of motion for the probability amplitudes of single and double excitation states. Following the description of single and double excitation systems, in section~\ref{sec3} we derive expressions for the concurrence between the atoms. We pay particular attention to the role of retardation in modifying the evolution of an initial entanglement. We show that the evolution of entangled states is affected drastically by the retardation effects that the harmonic oscillations of the concurrences characteristic of the single-mode interaction are suppressed in the presence of the retardation effects. At the same time we show that the evolution of the concurrence from an initial state in which atoms are entangled follows the evolution of the atomic population. The numerical results for various special cases of the time evolution of the concurrence are illustrated. Finally, in section~\ref{Conclusion} we summarize our results.

\section{The system and meaning of retardation}\label{sec1}

\begin{figure}[h]
\centering
\includegraphics[width=0.5\linewidth]{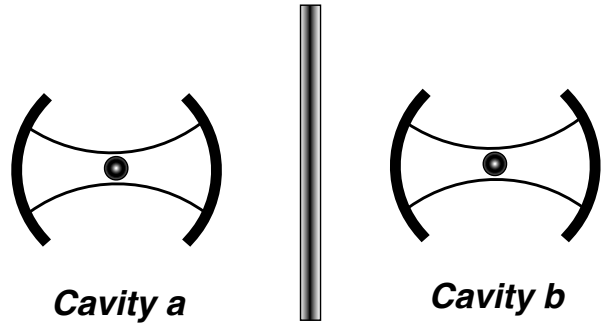}
\caption{(Color online) Schematic diagram of the system considered. The total system is composed of two independent Jaynes-Cummings systems, each containing a cavity and an atom located inside a one-dimensional, in general, multi-mode cavity field.}
\label{fig1}
\end{figure}

We consider a system composed of two independent Jaynes-Cummings (JC) cavities~\cite{JC63,pa63}. Each~JC system is composed of a one-dimensional cavity containing an atom located at a position $\vec{x}$ inside the cavity modes. The atoms are modelled as two-level systems with the excited $\ket{e_{i}}$ and ground $\ket{g_{i}}$ states separated by the transition frequency $\omega_a$, as illustrated in Fig.~\ref{fig1}.
The cavities are considered as composed of a multi-mode field with frequency difference between adjacent modes (free spectral range) such that multiple modes are supported within the atomic resonance line width. The need to consider a multi-mode rather than single-mode cavity results from the time-energy uncertainty relation that a finite interaction time between the atom and the cavity field leads to unavoidable spread of the achievable energy distribution.  

The interaction Hamiltonian between the atom and the multi-mode field, in the rotating-wave approximation, is of the form
\begin{equation}
H_{af} = i\hbar\sum_{n}\sum_{j=1}^{2}\left(g_{jn}(\vec{x}_j)a_{n}S^{+}_{j} - \rm{H.c.}\right) ,\label{eq8}
\end{equation}
where $g_{jn}(\vec{x}_j)$ is the position-dependent Rabi frequency which determines the strength of the coupling of the $j$th atom to the mode $n$ of the cavity field. The Rabi frequency is proportional to the density of the cavity modes to with the atom is coupled and can be written in a form
\begin{equation}
g_{jn}(\vec{x}_j) = g(\omega_{n})\left(\vec{d}_{j}\cdot
\hat{e}_{n}\right) {\rm e}^{i \vec{k}_{n}\cdot\vec{x}_j} ,\label{eq9}
\end{equation}
where $g(\omega_{n})$ characterise the density of the cavity modes of frequencies $\omega_{n}$ and $\hat{e}_{n}$ is the unit polarization vector of the mode $n$.

The function can be identified as the Airy function of the cavity so that the form of~$g(\omega_{n})$ depends on the type of the cavity. For a high-finesse optical cavity the function is of appreciable magnitude only at a single frequency $\omega_{c}$ called the cavity resonance. The function can be approximated by a delta function $\delta(\omega_{c})$ if the cavity is composed of perfectly reflecting mirrors, or by a Lorentzian of bandwidth $\kappa$, if the cavity losses through the mirrors with a rate $\kappa$ are included. For larger cavities (microwave cavities) with plane perfectly reflecting mirrors, the appropriate form of the function~$g(\omega_{n})$~is 
\begin{equation}
g(\omega_{n}) = \sqrt{\frac{\omega_{n}}{2\hbar\epsilon_0 L}} ,\label{eq9a}
\end{equation}
where $\omega_{n}=2 \pi n c/L$ is in the frequency of the modes set by the periodic boundary conditions of the cavity of length $L$. Similar results can be obtained for a confocal cavity where resonance frequencies of the transverse modes either overlap or fall exactly halfway between the longitudinal mode resonances~\cite{yariv}. If the linewidth $\Gamma$ of an atom located inside the cavity is much larger than the frequency difference between the modes, $\Gamma\gg c/L$, and the cavity bandwidth, $c/L\gg \kappa$, the atom may couple to $n$ discrete modes fitting inside the atomic linewidth and differing in frequency by $c/L$. In this case, the atom can be re-excited at finite time $L/c$ and also at multiples of $L/c$. In other words, multiple finite times are required for a photon emitted by the atom to cycle through the cavity and be reabsorbed by the atom. Thus, the retardation may result in the excitation arising from a finite time required for the photon to travel between the atom and the cavity mirrors.

\section{Evolution of entangled states}\label{sec2}

The main motivation of considering the multimode interaction between the atom and the cavity field is to study the drastic departure of the dynamics of entanglement from the usual single-mode behaviour~\cite{Eberly1,Eberly2,Eberly3,yelattice,sainz,Ficek1}. We consider situations when only single and double excitations are present in the system. In the case of single excitation, we assume that either the atoms or the cavity modes are initially prepared in a single excitation entangled state and discuss in details the dynamics of entanglement for two initial states
\begin{equation}
\ket{\psi(0)} = \left(\cos\theta\,\ket{e_{1}g_{2}} +\sin\theta\,\ket{g_{1}e_{2}}\right)\otimes\ket{\{0\}_\mu\{0\}_\nu} ,\label{eq24}
\end{equation}
which is a tensor product of a single excitation entangled state of the atoms and the vacuum states of both cavities, and
\begin{equation}
\ket{\psi(0)} = \left(\cos\theta\,\ket{\{1\}_{\mu}\{0\}_{\nu}} + \sin\theta\,\ket{\{0\}_{\mu}\{1\}_{\nu}}\right)\otimes\ket{g_{1}g_{2}} ,\label{eq25}
\end{equation}
which is a tensor product of a single excitation entangled state of the cavities and the atomic ground states. Here, $\theta$ determines the degree of entanglement in the initial state, and we use the index $\mu$ to label modes of the cavity $a$, and $\nu$ those of the cavity $b$.

In the case of double excitation, we choose an initial state of the form
\begin{eqnarray}
\ket{\phi(0)} = \left(\cos\theta\,\ket{g_1g_2}+ \sin\theta\,\ket{e_{1}e_{2}}\right)\otimes\ket{\{0\}_\mu\{0\}_\nu} ,\label{eq26}
\end{eqnarray}
which is a tensor product of a double excitation entangled state of the atoms and the vacuum states of both cavities.

\subsection{Evolution of single-excitation states}\label{case-single}

First, suppose that only a single excitation is present in the system. In this case, the Schr\"{o}dinger equation 
\begin{equation}
i\hbar\frac{\partial \ket{\psi(t)}}{\partial t} = H_{af}\ket{\psi(t)} ,\label{eq13}
\end{equation}
for the evolution of the state vector 
\begin{eqnarray}
\ket{\psi(t)} &=& C_{1}(t)\,\ket{e_{1} g_{2}\{0\}_{\mu}\{0\}_{\nu}} + C_{2}(t)\,\ket{g_{1} e_{2}\{0\}_{\mu}\{0\}_{\nu}}\nonumber\\
&+& \left(\sum_{\mu}C_{a\mu}(t)\,\ket{\{1\}_{\mu}\{0\}_{\nu}}
+\sum_{\nu}C_{b\nu}(t)\,\ket{\{0\}_{\mu}\{1\}_{\nu}}\right)\otimes\ket{g_{1}g_{2}} ,\label{eq27}
\end{eqnarray}
leads to a simple set of coupled differential equations for the probability amplitudes 
\begin{eqnarray}
\dot{C}_{1}(t) &=& \sum_{\mu}g_{1\mu}C_{a\mu}(t) , \label{eq28} \\
\dot{C}_{a\mu}(t) &=& -i\Delta_{\mu} C_{a\mu}(t) - g^{\ast}_{1\mu}C_{1}(t) ,\label{eq29}
\end{eqnarray}
and 
\begin{eqnarray}
\dot{C}_{2}(t) &=& \sum_{\nu}g_{2\nu} C_{b\nu}(t) , \label{eq30} \\
\dot{C}_{b\nu}(t) &=& -i\Delta_{\nu} C_{b\nu}(t) - g^{\ast}_{2\nu}C_{2}(t) ,\label{eq31}
\end{eqnarray}
where sums are taken over the discrete modes of the cavities, $\Delta_{\mu}=\omega_{\mu}-\omega_a$ and $\Delta_{\nu}=\omega_{\nu}-\omega_a$ are detunings of
the mode frequencies $\omega_{\mu}$ and $\omega_{\nu}$ of the cavity $a$ and $b$ from the atomic transition frequency $\omega_a$, respectively. For simplicity, we assume that $\omega_{a}$ coincides with the central mode frequency of each cavity. Also we note that the state (\ref{eq27}) involves superpositions of the cavity modes oscillating with amplitudes $C_{a\mu}(t)$ and $C_{b\nu}(t)$.

The formal integration of~(\ref{eq29}) and~(\ref{eq31}) yields 
\begin{eqnarray}
C_{a\mu}(t) = C_{a\mu}(0){\rm e}^{-i\Delta_{\mu}t} -
 \int_{0}^{t}dt^{\prime}g^{\ast}_{1\mu}C_{1}(t^{\prime}){\rm
e}^{-i\Delta_{\mu}(t-t^{\prime})}  ,\label{eq32}\\
C_{b\nu}(t) = C_{b\nu}(0){\rm e}^{-i\Delta_{\nu}t} -
 \int_{0}^{t}dt^{\prime}g^{\ast}_{2\nu}C_{2}(t^{\prime}){\rm
e}^{-i\Delta_{\nu}(t-t^{\prime})}  .\label{eq33}
\end{eqnarray}

On substituting for $C_{a\mu}(t)$ from (\ref{eq32}) in (\ref{eq28}), and for $C_{b\nu}(t)$ from (\ref{eq33}) in (\ref{eq30}), we arrive at the following two integro-differential equations
\begin{eqnarray}
\dot{C}_{1}(t) &=& \sum_{\mu}g_{1\mu} C_{a\mu}(0){\rm e}^{-i\Delta_{\mu}t} - \int_{0}^{t}dt^{\prime}
\sum_{\mu}|g_{1\mu}|^{2}C_{1}(t-t^{\prime})\,{\rm e}^{-i\Delta_{\mu}t^{\prime}} , \label{eq34} \\
\dot{C}_{2}(t) &=& \sum_{\nu}g_{2\nu} C_{b\nu}(0){\rm e}^{-i\Delta_{\nu}t} - \int_{0}^{t}dt^{\prime}
\sum_{\nu}|g_{2\nu}|^{2}C_{2}(t-t^{\prime})\,{\rm e}^{-i\Delta_{\nu}t^{\prime}} .\label{eq35}
\end{eqnarray}
The effect of retardation in the atom-field interaction is is clearly visible in terms of the detunings $\Delta_\mu$ and $\Delta_{\nu}$, which in general are all different, because the frequencies $\omega_{\mu}$ and $\omega_{\nu}$ are different. Since a large cavity of the length $L$ contains a finite number of discrete modes whose frequencies differ by $2 \pi c/L$, the detunings also differ by integer multiples of $2\pi c/L$. Thus, the dynamics of the system response will exhibit sharp peaks at discrete times due to the constructive interference of all modes.

\subsection{Evolution of double-excitation states}

We now turn to the problem of determining of how the retardation in the atom-field interaction can affect the evolution of double-excitation states. For this purpose we consider the situation when the excitations are evenly redistributed through the system, i.e., each of the Jaynes-Cummings cavities contains a single excitation. In this case, the state vector of the system can be written as
\begin{eqnarray}
\ket{\phi(t)} &=& D_{11}(t)|e_{1}e_{2}\{0\}_{\mu}\{0\}_{\nu}\rangle\!+\!\sum_{\nu}D_{2\nu}(t)|e_{1}g_{2}\{0\}_{\mu}\{1\}_{\nu}\rangle\!+\!\sum_{\mu}D_{3\mu}(t)\ket{g_{1}e_{2}\{1\}_{\mu}\{0\}_{\nu}} \nonumber\\
&+& \!\left(D_{00}(t)\ket{\{0\}_{\mu}\{0\}_{\nu}}
+\sum_{\mu}\sum_{\nu}D_{4\mu\nu}(t)\ket{\{1\}_\mu\{1\}_{\nu}}\right)\otimes\ket{g_{1}g_{2}} ,\label{eq37}
\end{eqnarray}
where $\{m\}_{\mu}\{n\}_{\nu}$ denotes the state of the cavity modes with $m$ photons in the mode $\mu$ of the cavity $a$, and $n$ photons in the mode $\nu$ of the cavity $b$. We have also included the auxiliary state $\ket{g_{1}g_{2}\{0\}_{\mu}\{0\}_{\nu}}$ with zero excitation that as one can see from (\ref{eq37}) creates  superpositions of the cavity modes crucial for the presence of an entanglement in the system.

On making use the Schr\"odinger equation, we arrive at the following set of coupled equations of motion for the probability amplitudes
\begin{eqnarray}
\dot{D}_{00}(t) &=& 0 ,\nonumber\\
\dot{D}_{11}(t) &=& -\sum_{\nu} g^{*}_{2\nu} D_{2\nu}(t)
-\sum_{\mu} g^{*}_{1\mu} D_{3\mu}(t) ,\nonumber\\
\dot{D}_{2\nu}(t) &=& -i\Delta_{\nu}D_{2\nu}(t) + g_{2\nu}\tilde{D}_{11}(t)
-\sum_{\mu} g^{*}_{1\mu} D_{4\mu\nu}(t) ,\nonumber\\
\dot{D}_{3\mu}(t) &=& -i\Delta_{\mu}D_{3\mu}(t) + g_{1\nu}D_{11}(t)
-\sum_{\nu} g^{*}_{2\nu}D_{4\mu\nu}(t) ,\nonumber\\
\dot{D}_{4\mu\nu}(t &=& -i(\Delta_{\mu} + \Delta_{\nu})D_{4\mu\nu}(t) 
+ g_{1\mu}D_{2\nu}(t)+g_{2\nu}D_{3\mu}(t) .\label{eq38}
\end{eqnarray}
The equations (\ref{eq38}) are cumbersome because the involvement of many modes of the cavities. Therefore, we will use numerical methods to solve them for a given initial condition.

\section{Evolution of entanglement}\label{sec3}

Since we are interested in the dynamics of entanglement between the atoms and also between the cavity modes that behaviour like two-qubit systems, we shall consider concurrence as a measure of entanglement. The concurrence is defined as~\cite{W98}
\begin{equation}
\mathcal{C} = \rm{max}\{0,\sqrt{\lambda_1}-\sqrt{\lambda_2}-\sqrt{\lambda_3}-\sqrt{\lambda_4}\} ,\label{eq10}
\end{equation}
where $\lambda_i$ are the eigenvalues (in descending order) of the Hermitian matrix $R=\rho\tilde{\rho}$, in which~$\rho$ is the density matrix of the system, $\tilde{\rho} = \sigma_y\otimes\sigma_{y}\,\rho^{\ast}\sigma_y\otimes\sigma_y$ and $\sigma_y$ is a Pauli matrix. The concurrence varies between $0$ when qubits are separable and $1$ when they are maximally entangled. 

In the case of a single excitation present in the system, the reduced density operator of the atoms $\rho_{AB}={\rm Tr}_{ab}(\rho)$ can be written in the basis of the product states, $|1\rangle=|e_{1}e_{2}\rangle,|2\rangle=|e_{1}g_{2}\rangle,|3\rangle=|g_{1}e_{2}\rangle,|4\rangle=|g_{1}g_{2}\rangle$, in which it has a form 
\begin{eqnarray}
\rho_{AB} &=& \left(
\begin{array}{cccc}
0 & 0 & 0 & 0\\
0 &|C_{1}(t)|^{2} & C^{\ast}_{1}(t)C_{2}(t) & 0\\
0 & C_{1}(t)C^{\ast}_{2}(t) & |C_{2}(t)|^{2} & 0\\
0 & 0 & 0 & \sum_{\mu}|C_{a\mu}(t)|^{2}+\sum_{\nu}|C_{b\nu}(t)|^{2}
\end{array}\right) ,\label{eq11}
\end{eqnarray}
from which we can easy find the concurrence 
\begin{equation}
\mathcal{C}_{AB}(t) = 2\left|C_{1}(t)C_{2}^{\ast}(t)\right| = 2|\rho_{23}(t)| .\label{eq12}
\end{equation}
Thus, the necessary and sufficient condition for entanglement between the atoms is the present of the one-photon coherence~$|\rho_{23}(t)|$. 

In the case of a double excitation, again by tracing over the cavity modes, the reduced density operator of the atoms written in the basis of the product states states takes a matrix form
\begin{eqnarray}
\rho_{AB} &=& \left(\!
\begin{array}{cccc}
|D_{11}(t)|^{2} & 0 & 0 & D_{11}(t)D_{00}^{*}(t)\\
0 &\sum_{\nu}\!|D_{2\nu}(t)|^{2} & 0 & 0\\
0 & 0 & \sum_{\mu}\!|D_{3\mu}(t)|^{2} & 0\\
D_{11}^{*}(t)D_{00}(t) & 0 & 0 &
|D_{00}(t)|^{2}\!+\!\sum_{\nu\mu}\!|D_{4\mu\nu}(t)|^{2}
\end{array}\!\right) ,\label{eq42}
\end{eqnarray}
from which it is straightforward to evaluate the concurrence
\begin{eqnarray}
\mathcal{C}_{AB}(t) = 2\, {\rm max}\left\{0,\tilde{\mathcal{C}}_{AB}(t)\right\} ,\label{eq43}
\end{eqnarray}
where
\begin{eqnarray}
\tilde{\mathcal{C}}_{AB}(t) = |D_{11}(t)||D^*_{00}(t)| -\sqrt{\left(\sum_{\nu}|D_{2\nu}|^2\right)\left(\sum_{\mu}|D_{3\mu}|^2\right)} .\label{eq44}
\end{eqnarray}
The entangled properties of the system are clearly exhibited by the off-diagonal elements of the density matrix (\ref{eq42}). It is apparent from (\ref{eq44}) that the presence of the auxiliary state with zero excitation, $D_{00}(t)\neq 0$, is necessary to produce entanglement between the atoms.

\subsection{The case of single-mode cavities}

\begin{figure}[h]
\center{\includegraphics[scale=0.3]{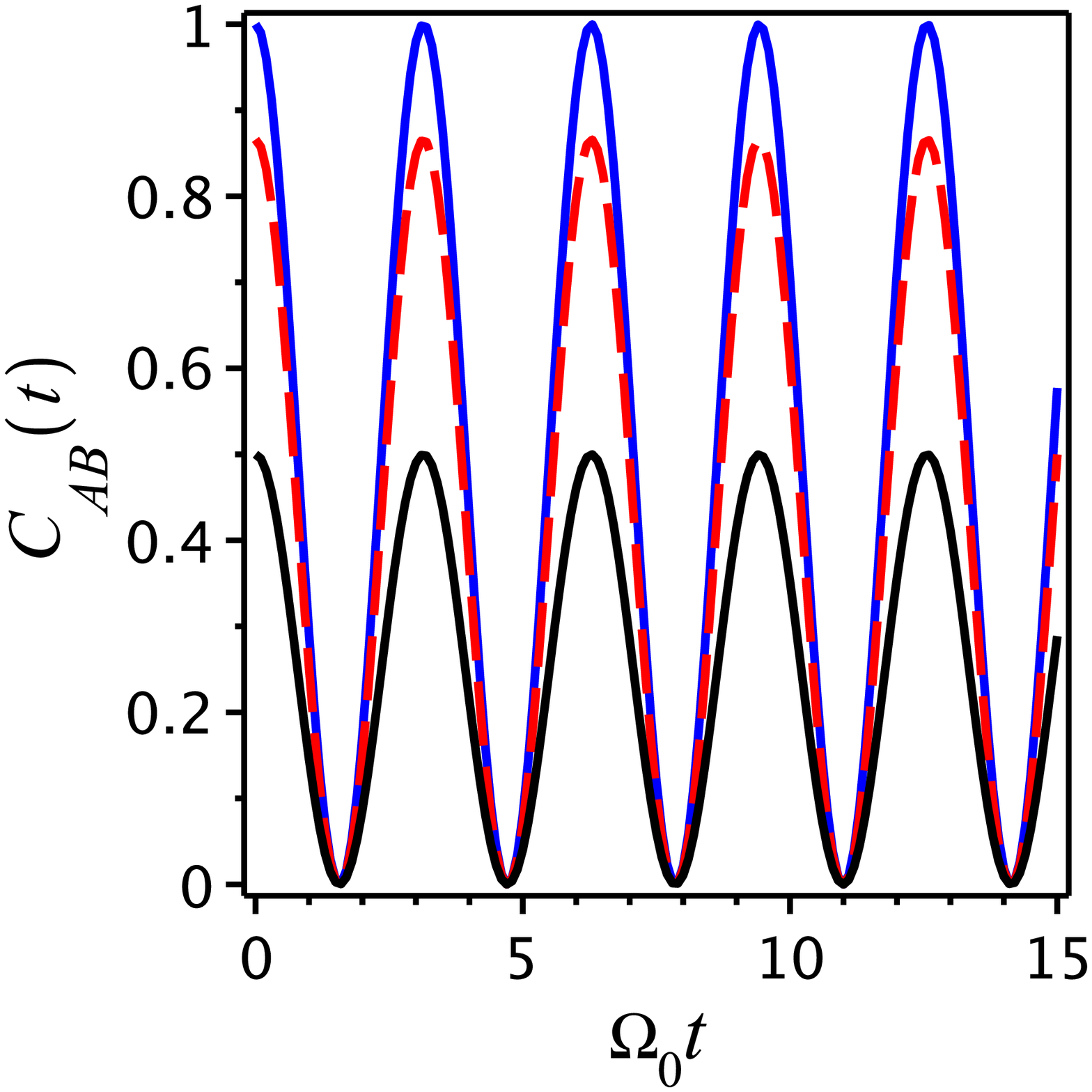}\hspace{0.5cm}\includegraphics[scale=0.3]{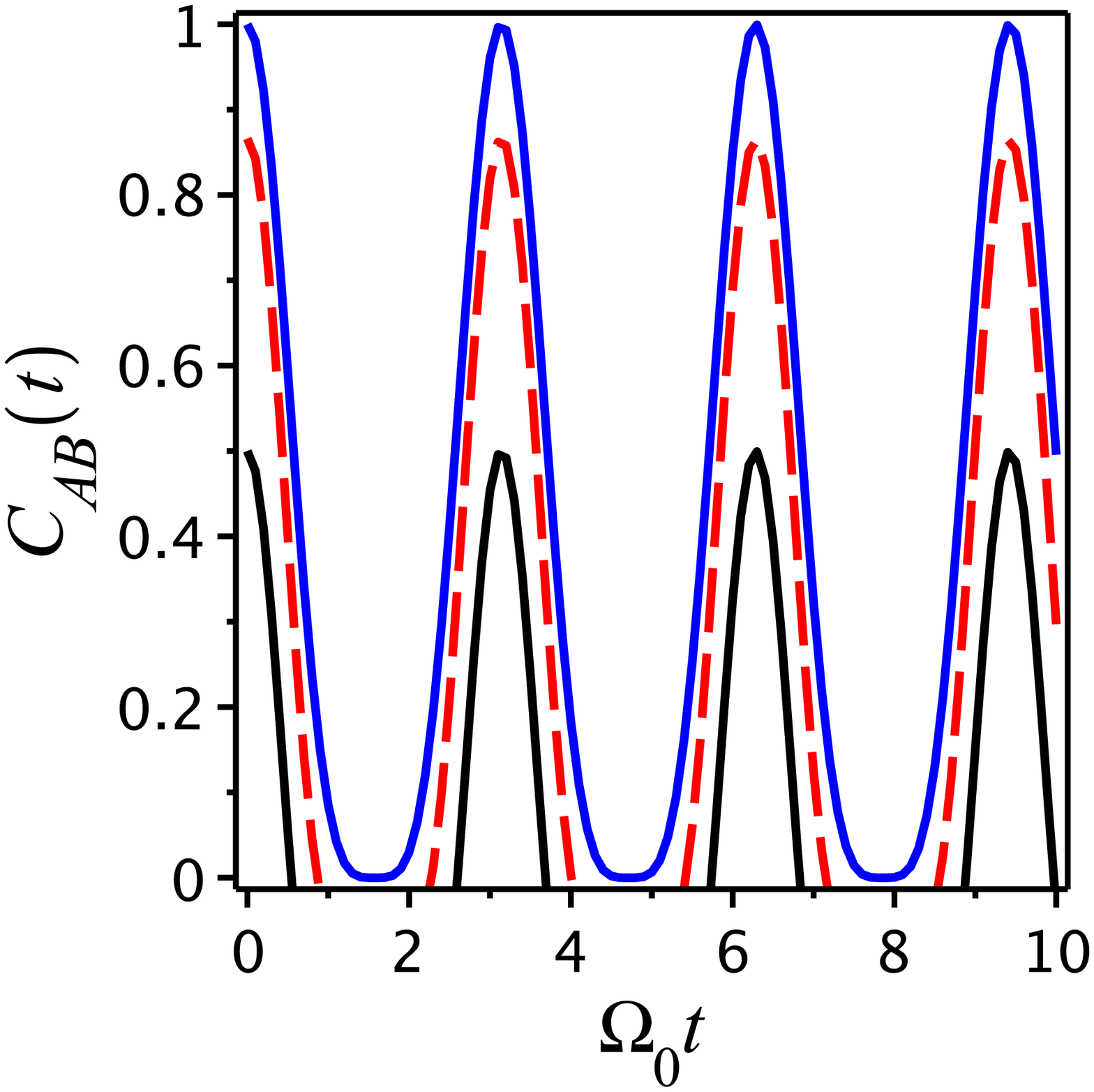}}
\caption{(Color online) The concurrences $\mathcal{C}_{AB}(t)$ plotted as a function of dimensionless time $\Omega_{0}t$ for the initial single-excitation state (\ref{eq24}) (left frame) and the double-excitation state (\ref{eq26}) (right frame) with $\theta=\pi/4$ (solid blue line), $\theta=\pi/6$ (dashed red line), and $\theta=\pi/12$ (solid black line).}
\label{fig2}
\end{figure}

Let us first present the results for the evolution of entanglement in the simplified double Jaynes-Cummings system composed of two single-mode $(\mu =\nu =1)$ cavities~\cite{Eberly1,Eberly2,Eberly3,yelattice,sainz,Ficek1}.
Figure~\ref{fig2} shows the time evolution of the concurrences for several different values of the parameter $\theta$. It is seen that an initial entanglement periodically oscillates in time and vanishes at certain discrete times $t_{n} = n\pi/2,\, (n=1,3,5,\ldots)$ when it is completely transferred to the cavity modes. This behaviour is seen for both maximally $(\theta =\pi/4)$ and non-maximally entangled states $(\theta\neq \pi/4)$, but the effect of going off maximally entangled initial states is clearly to decrease the amount of entanglement~\cite{Meyster1,Eberly1,Eberly2,Eberly3,Ficek1}. 

Similar behaviour of the periodic exchange of entanglement is observed for double excitation states with a difference that in this case the phenomenon of the sudden death of entanglement is observed. Therefore, we may conclude that in the case of single-mode cavities the periodic oscillatory behaviour of the concurrence can be observed as resulting from an instantaneous exchange of an excitation between the atoms and the cavity modes. This is true, however, only for sufficiently small cavities in which a light signal can pass between the atom and the cavity boundaries in a time short compared with the time $\Delta t$ required for appreciate (secular) changes in the atomic states. In other words, the length $L$ of the cavity must be much smaller than $c\Delta t$.

\subsection{The case of multi-mode cavities}

\begin{figure}[h]
\center{\includegraphics[scale=0.3]{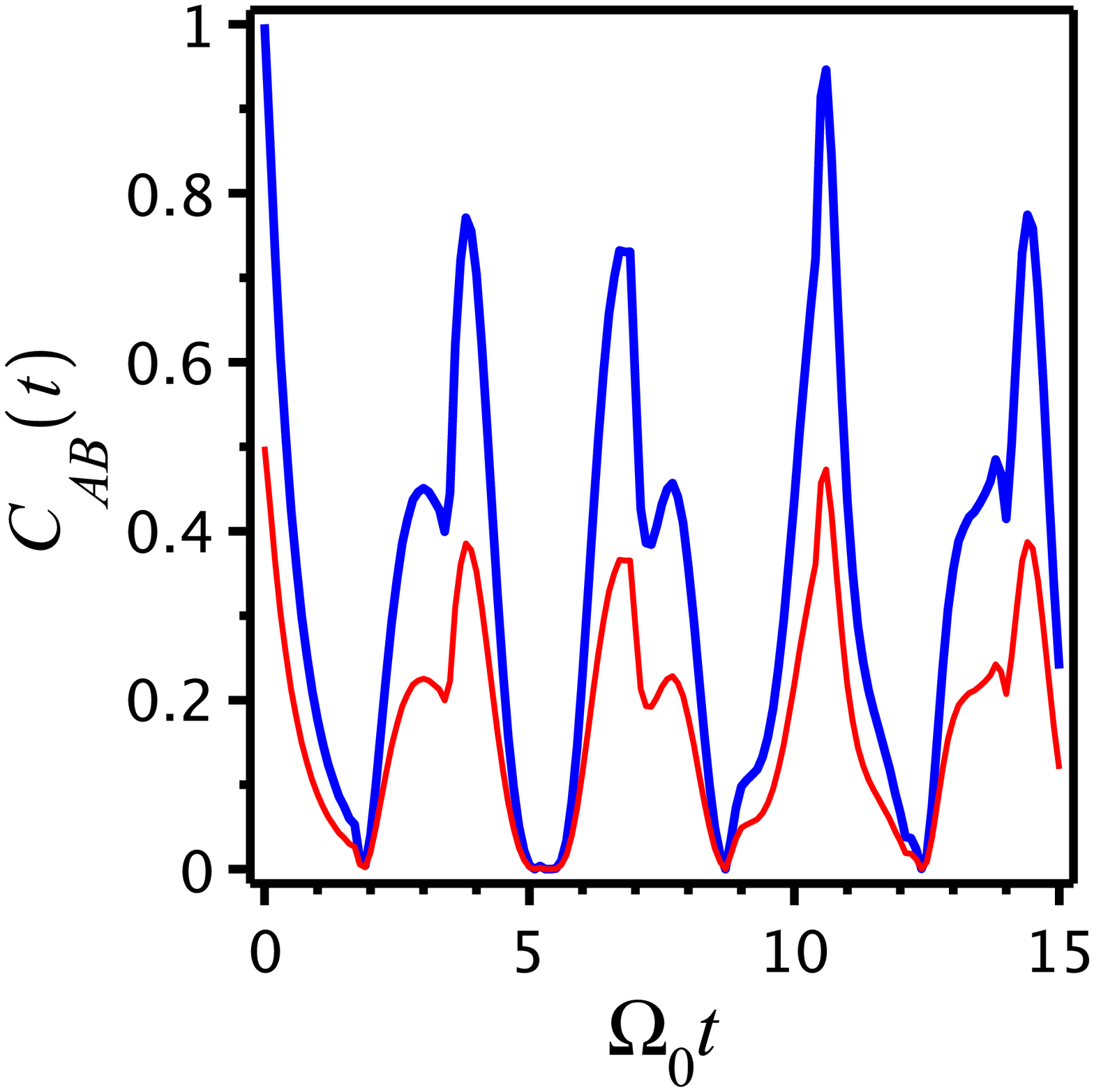}\hspace{0.5cm}\includegraphics[scale=0.3]{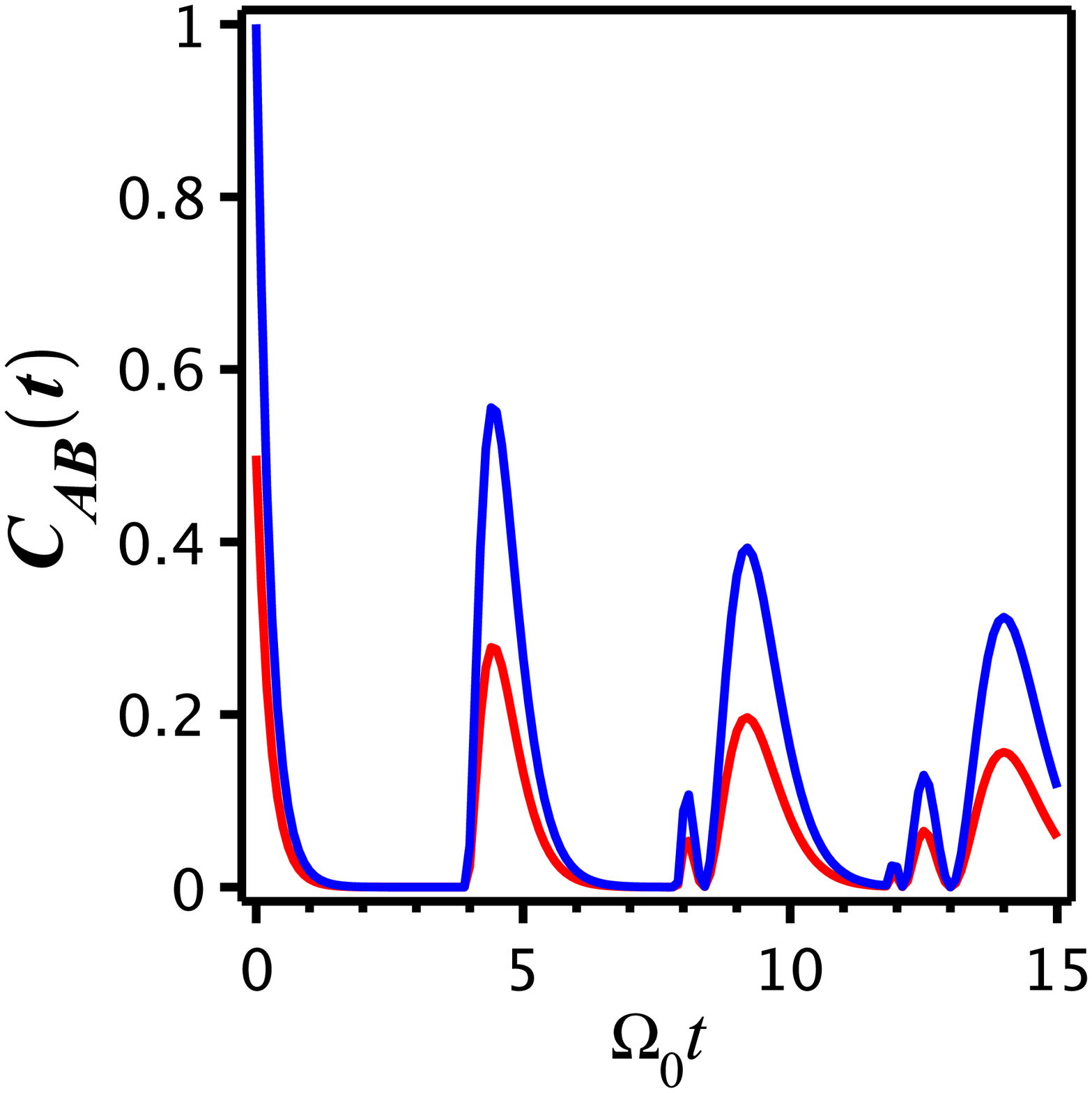}}
\caption{(Color online) The time evolution of $\mathcal{C}_{AB}(t)$ for the initial state (\ref{eq24}) with $\theta=\pi/4$ (solid blue line) and $\theta=\pi/12$ (solid red line) and the number of the cavity modes $n=19$ (left frame) corresponding to $L=6.7\times 10^{2}\lambda_{a}$ and $n=99$ (right frame) corresponding to $L=3.48\times 10^{3}\lambda_{a}$. The other parameters are $\omega_a=4.84\times 10^{3}\Omega_{0}$, where $\lambda_{a}$ is the wavelength of the atomic transition and $\Omega_{0}$ is the vacuum Rabi frequency of the central cavity mode.}
\label{fig4}
\end{figure}

\begin{figure}[h]
\center{\includegraphics[width=0.5\linewidth ]{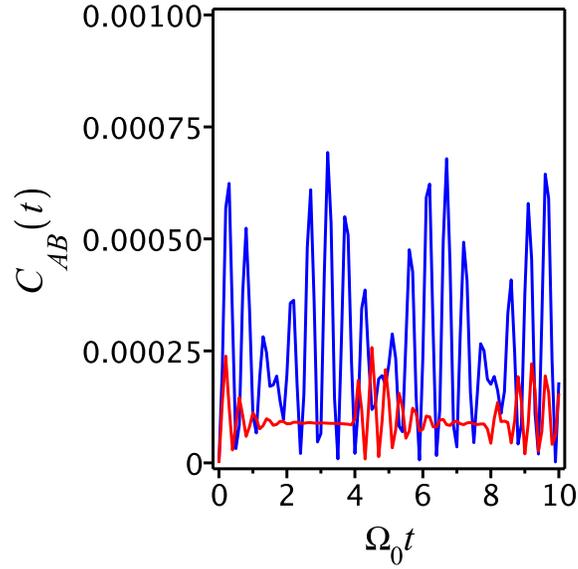}}
\caption{(Color online) The time evolution of the concurrence $\mathcal{C}_{AB}(t)$ for $\omega_a=1.11\times 10^4\Omega_{0}$ and different number of modes: $n=19$ (solid blue line) corresponding to $L=6.7\times 10^{2}\lambda_{a}$ and $n=99$ (solid red line) corresponding to $L=3.48\times 10^{3}\lambda_{a}$. The system was initially in the state~(\ref{eq25}) with $\theta = \pi/4$.}
\label{fig5}
\end{figure}

We now examine the effect of retardation on the dynamics of entanglement for different numbers of modes and for two different initial states. Figure~\ref{fig4} shows the time evolution of the concurrence for the initial state (\ref{eq24}) in which the atoms are initially entangled, whereas in Fig.~\ref{fig5} the initial state is (\ref{eq25}) in which the cavity modes are initially entangled. The concurrence is determined from~(\ref{eq12}), where the time evolution of the probability amplitudes is found by solving numerically the sets of coupled differential equations~(\ref{eq34}) and (\ref{eq35}). 

The evolution of the concurrence in the presence of the retardation is seen to be qualitatively different from the previous case of the instantaneous exchange of an excitation between the atoms and the cavity modes, shown in Fig.~\ref{fig2}. Firstly, the evolution is not oscillatory with time, and secondly the initial entanglement is not preserved in time, it decreases as time progresses. At the initial time, the concurrence decays almost exponentially to zero and remains zero until a finite time at which a nonzero entanglement suddenly revivals. The process repeats with increasing distortions of the concurrence. The effect of increasing number of modes to which the atom couples is to decrease the periodicity of the entanglement revival. In other words, the process of revival of the entanglement degrades with an increasing number of modes to which the atoms are coupled. Equivalently, we may say that the memory effects of the atom-field interaction degrade with an increasing number of the field modes. 

Suppose now that the system starts at $t=0$ from the state (\ref{eq25}) in which the cavity modes are entangled and the atoms are in their ground states. In this case, the transfer of the initial entanglement may depend strongly on the number of the cavity modes to which the atoms are coupled. Figure~\ref{fig5} shows the time evolution of the concurrence between the atoms for two different numbers of cavity modes. The effect of increasing the number of modes is seen to decrease the efficiency of transferring entanglement from the field to the atoms, namely, the entanglement between the atoms decreases in magnitude and occurs in more restricted intervals of time. This is readily understood if it is recalled that increasing the number of modes leads to a smaller efficiency of transferring an initial entanglement existing between two modes of the cavities to the atoms. One may notice that the concurrence exhibits collapses and revivals on time scales much larger than the Rabi period, the phenomenon characteristics of the Jaynes-Cummings system~\cite{SM83}. 

\begin{figure}[h]
\center{\includegraphics[scale=0.3]{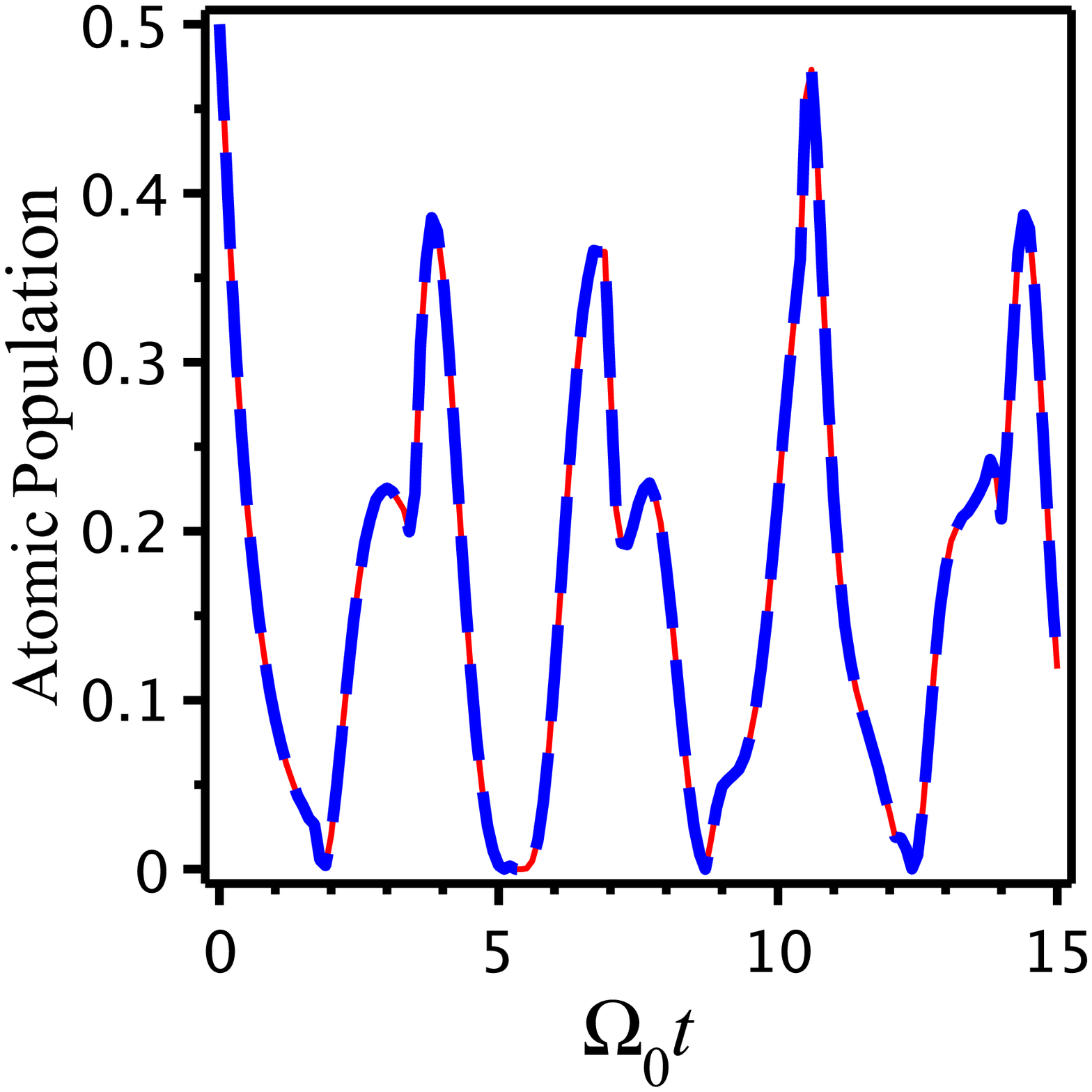}\hspace{0.5cm}\includegraphics[scale=0.3]{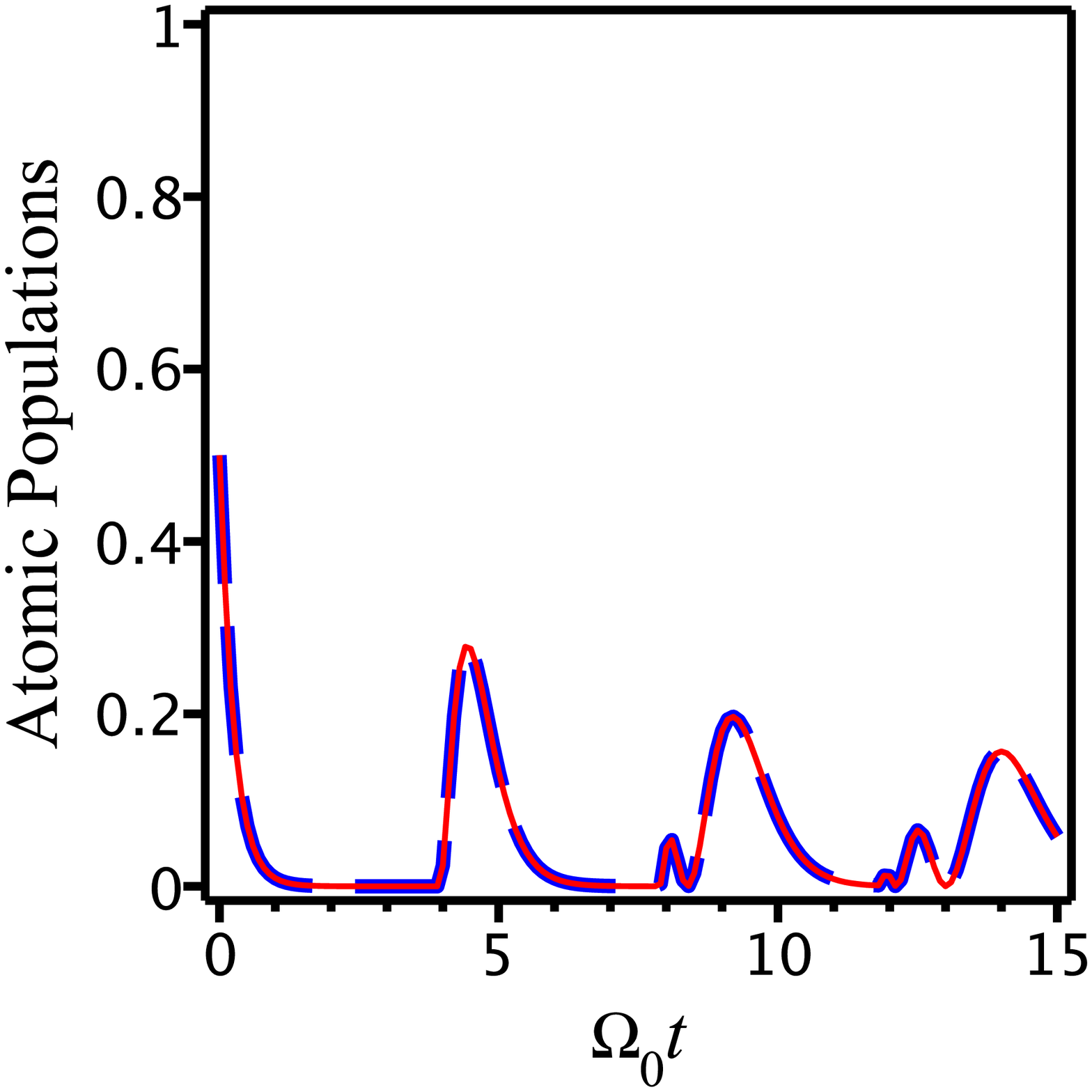}}
\caption{(Color online) The time evolution of the probabilities $|C_{1}(t)|^2$ (solid red line) and $|C_{2}(t)|^2$ (dashed blue line) for $n=19$ (left frame) and $n=99$ (right frame). The system was initially in the state~(\ref{eq24}) with $\theta=\pi/4$.}
\label{fig6}
\end{figure}

It is interesting to note that the evolution of the concurrence from an initial state in which atoms are entangled follows the evolution of the atomic population. To see it, we plot in Fig.~\ref{fig6} the time evolution the probabilities $|C_{1}(t)|^2$ and $|C_{2}(t)|^2$ for the same parameters as in Fig.~\ref{fig4}. Clearly, the initially excited atoms decay almost exponentially in time~\cite{Meyster1}. However, at the particular times that correspond to $nL/c$, where $n$ is an integer, a sudden change (jump) in the probabilities occurs. These are just the times when the radiation field emitted into the cavity modes returns to the atoms. Correspondingly, the entanglement revivals and remains nonzero for a finite time the excitation resides in the atoms. 

The sudden jumps continue in time. However, the periodic maxima of the  populations are reduced in magnitude and become more broadened as~$t$ increases. This effect can be explained using the energy-time uncertainty arguments and is readily understood if it is recalled that the excitation wave packet spreads during the evolution, that the excitation becomes less localized as time progresses. This delocalisation of the excitation, it turns out, is sufficient to wipe out the entanglement. 
\begin{figure}[h]
\center{\includegraphics[width=0.55\linewidth ]{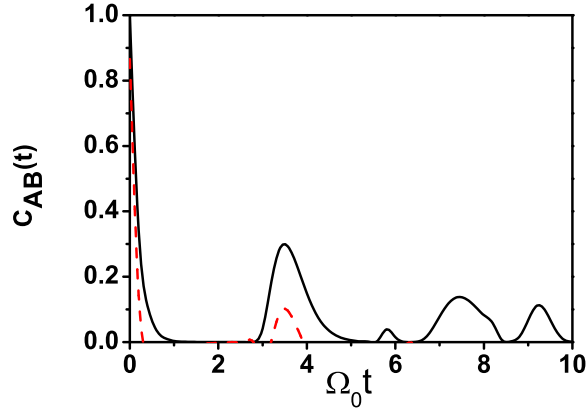}}
\caption{(Color online) The time evolution of the concurrence $\mathcal{C}_{AB}(t)$ for $n=49$ corresponding to $L=1.72\times 10^3\lambda_a$ and $\omega_a=1.11\times 10^4\Omega_{0}$. The system was initially in the state (\ref{eq26}) with $\theta=\pi/4$ (solid black line) and $\theta=\pi/6$ (dashed red line).}
\label{fig7}
\end{figure}

Figure~\ref{fig7} shows the time evolution of the concurrence for two initial conditions of maximally and non-maximally entangled doubly excited state (\ref{eq26}). At very early times, the concurrence decays almost exponentially to zero, remains zero for a finite period of time and then revivals. The concurrence remains nonzero for a finite time, and the process repeats with increasing distortion. The distortion is significantly much stronger than that observed previously for the evolution of the single excitation states.
The reason is that the evolution of the excitations occurs in two JC cavities with no fixed relation between them. At short times the excitations in the two JC cavities are unresolved resulting in the revival of the entanglement, but after a long time, the excitations can be resolved which gives no inkling of correlations between the atoms.

\section{Summary}\label{Conclusion}

We have studied the effect of retardation on the evolution of entanglement in a system composed of two independent Jaynes-Cummings cavities. The retardation effects could be important in large cavities where the mode spacing is small enough such that an atom located inside the cavity can concurrently couple to many modes of the cavity field. We have used the concurrence to quantify entanglement between the atoms and the cavity modes and have shown that the multi-mode structure of the field inside the cavity affects considerably the harmonic evolution of the atomic population and entanglement characteristic of single-mode Jaynes-Cummings systems. We have considered both single and double excitation cases and have shown that the retardation leads to sudden death of the initial entanglement and abrupt revival at times corresponding to intervals required for a round trip of the excitation in the cavity. This demonstrates that in the presence of the retardation, the complete transfer of entanglement during the evolution of the independent JC systems never occurs, which is the essential feature of single mode Jaynes-Cummings systems. Finally, we have shown that single-excitation entangled states are more robust against the retardation than doubly-excited~states. 

\section*{Acknowledgments}

This work was supported by a research grant from the King Abdulaziz City for Science and Technology (KACST).

\end{document}